\documentclass[pra,aps,amsmath,amssymb,reprint,superscriptaddress, floatfix]{revtex4-2}

\usepackage{graphicx}% Include figure files
\usepackage{dcolumn}% Align table columns on the decimal point
\usepackage{bm}% bold math
\usepackage{ulem} % strikethrough
\usepackage{siunitx} % Units, micrometer in particular
\usepackage{caption}
\usepackage{braket}
\usepackage{xcolor}
\usepackage[colorlinks,bookmarks=false,
linkcolor=blue,
urlcolor=blue,
citecolor=blue
]{hyperref}
\usepackage[newcommands]{ragged2e}
\usepackage[round-mode=places]{siunitx}
\begin{document}
% \bibliographyunit[\chapter]

\preprint{APS/123-QED}

% \title{Flux Modulated Tunable Couplings of Two Strongly Nonlinear Oscillators}

% \title{Flux Modulated Tunable Couplings and Level Attraction of Two Strongly Nonlinear Oscillators}

\title{Flux-modulated tunable interaction regimes in two strongly nonlinear oscillators}

\author{J.~D. Koenig}
\email{jdkoenig23@gmail.com}
\affiliation{Kavli Institute of Nanoscience, Delft University of Technology, PO Box 5046, 2600 GA Delft, The Netherlands}

\author{G. Barbieri}
% \email{b@email2}
\affiliation{Kavli Institute of Nanoscience, Delft University of Technology, PO Box 5046, 2600 GA Delft, The Netherlands}
% \affiliation{affil2}

\author{F. Fani Sani}
% \email{b@email2}
\affiliation{Kavli Institute of Nanoscience, Delft University of Technology, PO Box 5046, 2600 GA Delft, The Netherlands}
% \affiliation{affil3}

\author{C.~A. Potts}
% \email{b@email2}
\affiliation{Kavli Institute of Nanoscience, Delft University of Technology, PO Box 5046, 2600 GA Delft, The Netherlands}
\affiliation{Niels Bohr Institute, University of Copenhagen, Blegdamsvej 17, 2100 Copenhagen, Denmark}
\affiliation{NNF Quantum Computing Programme, Niels Bohr Institute, University of Copenhagen, Denmark}

\author{M. Kounalakis}
\email{marios.kounalakis@gmail.com}
\affiliation{Kavli Institute of Nanoscience, Delft University of Technology, PO Box 5046, 2600 GA Delft, The Netherlands}
\affiliation{Luxembourg Institute of Science and Technology (LIST), 4362, Esch-sur-Alzette, Luxembourg}

\author{G.~A. Steele}
\email{g.a.steele@tudelft.nl}
\affiliation{Kavli Institute of Nanoscience, Delft University of Technology, PO Box 5046, 2600 GA Delft, The Netherlands}

% \affiliation{Kavli Institute of Nanoscience, Delft University of Technology, Lorentzweg 1, 2628 CJ, Delft, The Netherlands}

\date{\today}

\begin{abstract}
The ability to efficiently simulate a variety of interacting quantum systems on a single device is an overarching goal for digital and analog quantum simulators.
In circuit quantum electrodynamical systems, strongly nonlinear superconducting oscillators are typically realized using transmon qubits, featuring a wide range of tunable couplings that are mainly achieved via flux-dependent inductive elements.
Such controllability is highly desirable both for digital quantum information processing and for analog quantum simulations of various physical phenomena, such as arbitrary spin-spin interactions.
Furthermore, broad tunability facilitates the study of driven-dissipative oscillator dynamics in previously unexplored parameter regimes.
In this work, we demonstrate the ability to selectively activate different dynamical regimes between two strongly nonlinear oscillators using parametric modulation.
In particular, our scheme enables access to regimes that are dominated by photon-hopping, two-mode squeezing, or cross-Kerr interactions.
Finally, we observe level repulsion and attraction between Kerr-nonlinear oscillators in regimes where the nonlinearities exceed the coupling strengths and decay rates of the system.
Our results could be used for realizing purely analog quantum simulators to study arbitrary spin systems as well as for exploring strongly nonlinear oscillator dynamics in previously unexplored interaction regimes.
\end{abstract}

\maketitle

%\tableofcontents 

\section{Introduction}

Quantum information processors based on superconducting circuits have long relied on the transmon qubit as a robust, reliable, and high-coherence building block in the journey toward large-scale digital quantum computation \cite{Koch2007, OMalley2016, Arute2019}.
Circuit quantum electrodynamical (cQED) devices are also of great interest to the development of analog quantum simulators, in which devices are custom-built to emulate the behavior of distinct systems which are otherwise typically challenging to control or probe directly \cite{Georgescu2014, Hartmann2016}. Such devices may enable the probing of physics in otherwise inaccessible parameter regimes due to the high degree of engineerability in superconducting circuits enabled by modern nanofabrication techniques and materials science \cite{Mostame2016, Wilkinson2020}.
% \cite{ , Ganzhorn2020}.

Transmon-based cQED systems may be described as collections of coupled Kerr-nonlinear oscillators, which in recent years have been imbued with in-situ tunable resonance frequencies, couplings, and nonlinearities achievable by external control~\cite{Niskanen2006, Chen2014, Kounalakis2018, Yan2018}. Tunable couplers have been successfully used to implement high-fidelity two-qubit gates, enter novel coupling regimes, and are useful elements for mitigating undesirable interactions in designs for scalable quantum computing architectures~\cite{Geller2015, McKay2016, Roth2017, Collodo2020, Miyanaga2021, Stehlik2021, Sung2021, Heunisch2023, Janzen2023, Glaser2023, Li2024}. While such developments have contributed significantly to progress in digital gate-based architectures, there is still underexplored territory in using such platforms to emulate other interactions and physical systems such as extended Bose-Hubbard, arbitrary spin-spin, fractional Bloch oscillations, and lattice gauge theories \cite{Corrielli2013, Jin2013, Jin2013_2, Marcos2013, Marcos2014, Dekharghani2017, Sameti2017, Sameti2019, Rosen2024}.
Systems of coupled Kerr-nonlinear oscillators (KNOs) can be engineered to enable a variety of novel interactions, including intrinsic longitudinal and radiation-pressure-like couplings \cite{Didier2015, Rodrigues2021, Potts2024}. By introducing nonlinear coupling elements, many such interactions can be activated when driven or parametrically modulated~\cite{Sameti2019, Kounalakis2022, Kounalakis2023}. Moreover, control over all $\sigma_X \sigma_X$, $\sigma_Y \sigma_Y$, and $\sigma_Z \sigma_Z$ couplings individually would allow for analog simulation of arbitrary XYZ spin-model Hamiltonians and coupled Ising spins~\cite{Salathe2015, Puri2017, Sameti2019}. Devices with couplers containing more highly nonlinear elements may also be used to enter into regimes where strictly nonlinear couplings such as correlated photon hopping and photon-pair tunnelling terms dominate, allowing for the simulation of more exotic physics~\cite{Frattini2017}.

Here, we implement such flux-tunable interactions on a superconducting circuit containing two flux-tunable transmon qubits connected by a fixed capacitive coupling and a tunable nonlinear inductive coupling.
The latter is realized using a superconducting quantum interference device (SQUID). By parametrically modulating the external flux threading the SQUID loop of the coupler, we operate the device in regimes where the longitudinal (cross-Kerr) coupling is dominant over a two-mode squeezing interaction and in which the single-photon exchange interaction (beam-splitter) and cross-Kerr strengths are comparable. 
We observe two-mode squeezing effects through the use of parametric modulation, which, together with the single-photon hopping interaction, are characterized by level attraction and repulsion between the oscillators. While previous studies have explored such effects in linear systems~\cite{Bernier2018, Wang2019_2, Yu2019}, our system extends level attraction phenomena in KNOs. Our results pave the way for the realization of novel analog quantum simulators, based on nonlinear oscillators containing parametrically driven tunable couplers, to study exotic parameter regimes in nonlinear quantum systems.

\section{Theoretical analysis}
\begin{figure}[t]
    %\captionsetup{justification=raggedright}
    \captionsetup{justification = Justified,
              width=\linewidth}
    \includegraphics[width=\linewidth]{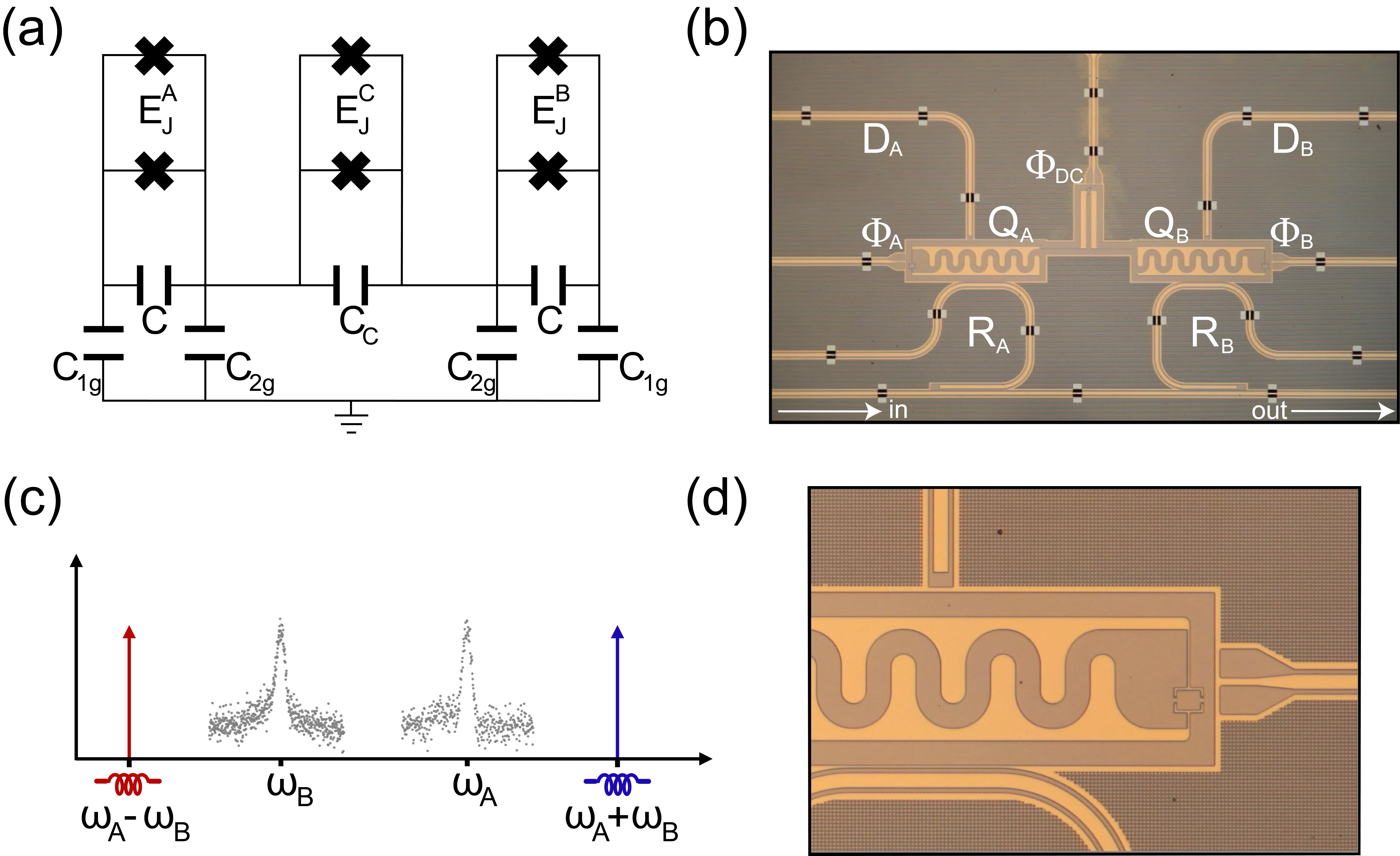}
    \caption{Device and measurement scheme. (a) Circuit diagram for the device. On the left and right sides are two flux-tunable transmon qubits consisting of SQUIDs with gate capacitances to ground. The tunable coupler in the center consists of a coupling capacitor and a symmetric SQUID. (b) Optical microscope image of the device, including transmission line, readout resonators ($R_i$), drive lines ($D_i$), flux lines ($\Phi_i$), two transmons ($Q_i$), and the tunable coupler. (c) Schematic of the experiment. The flux incident on the coupler SQUID loop is modulated at either the difference or sum frequency of the two transmons. (d) Optical microscope image of Qubit B.}
    \label{Circuit Diagram}
\end{figure}

The device comprised of two transmon qubits and a tunable coupler is shown in Fig.~\ref{Circuit Diagram}. The coupling between the two transmons is solely characterized by the charging and Josephson energies of the constituent circuit elements.
Both the linear and nonlinear interactions can be tuned via the total flux threading the coupler SQUID loop. The Josephson energy of the coupler is written as
\begin{multline}
E_J^C(\Phi_{DC})=E_{J \rm max}^C|\cos{\bigg(\pi\frac{\Phi_{DC}}{\Phi_0}\bigg)}|\\
\times \sqrt{1+d_C^2\tan^2{\bigg(\pi\frac{\Phi_{DC}}{\Phi_0}\bigg)}}
\label{ejc}
\end{multline}
where $E_{J \rm max}^C$ is determined by the inductance of the unbiased SQUID loop, $d_C$ is a measure of the asymmetry of the junction inductances comprising the SQUID, and $\Phi_{DC}$ is the DC flux threading the loop \cite{Koch2007, Hutchings2017}. Each of the two transmons is capacitively coupled to its own coplanar waveguide readout resonator, which are in turn coupled to a common feedline through which the device is driven and probed. The Josephson energies of the two qubits are related to their own flux biases $\Phi_A$ and $\Phi_B$ in the same form as Eq.~\ref{ejc}. In the coupled system, the ground to excited state transition frequency for transmon $\it{i}$ is given in units where $\, \hbar=1$ as
\begin{equation}
\begin{aligned}
\omega_i\approx\sqrt{8\widetilde{E}_J^iE_{C}^i}-E_{C}^i
\label{qubit freq}
\end{aligned}
\end{equation}
with $\widetilde{E}_J^i={E}_J^i+{E}_J^C/4$ the modified Josephson energy due to the coupler and $E_{C}^i$ the charging energy of transmon \textit{i}.

We obtain a full system Hamiltonian following the procedure outlined in the supplemental material and Ref.~\cite{Kounalakis2018}. We quantize the circuit shown in Fig.~\ref{Circuit Diagram}(a) and retain terms in the expansions of the energy potentials of the SQUIDs to fourth-order. We can obtain a simplified Hamiltonian given by
\begin{multline}
H=\omega_A\hat{a}^{\dagger}\hat{a}+\frac{\alpha_A}{2}\hat{a}^{\dagger}\hat{a}^{\dagger}\hat{a}\hat{a}
+\omega_B\hat{b}^{\dagger}\hat{b}+\frac{\alpha_B}{2}\hat{b}^{\dagger}\hat{b}^{\dagger}\hat{b}\hat{b}\\
+J_1(\hat{a}^{\dagger}\hat{b}+\hat{a}\hat{b}^{\dagger})+J_2(\hat{a}^{\dagger}\hat{b}^{\dagger}+\hat{a}\hat{b})+V\hat{a}^{\dagger}\hat{a}\hat{b}^{\dagger}\hat{b}
\label{base Hamiltonian}
\end{multline}
where we have defined:
\begin{multline}
    J_{1,2}:=
    \pm\left(\frac{\widetilde{E}_J^A\widetilde{E}_J^BE_{C}^AE_{C}^B}{4}\right)^{1/4}\\
    \times\left(\frac{E_C^C}{\sqrt{E_{C}^AE_{C}^B}}\mp\frac{E_J^C}{2\sqrt{\widetilde{E}_J^A\widetilde{E}_J^B}}\right)
    \label{J12}
\end{multline}
and
\begin{equation}
\begin{aligned}
    V:=-\frac{E_J^C}{8}\sqrt\frac{E_{C}^AE_{C}^B}{{\widetilde{E}_J^A\widetilde{E}_J^B}}
\end{aligned}
\end{equation}
where $J_1$ is the strength of the hopping interaction, $J_2$ the squeezing, $V$ the cross-Kerr, $\alpha_i\approx-E_{C}^i$ the anharmonicity, and we have neglected higher order terms which are far off-resonant from the relevant dynamics of the measurements performed. In systems of coupled Kerr-nonlinear oscillators, the two-mode squeezing term has been previously shown to produce nondegenerate parametric oscillations \cite{Bengtsson2018, Wustmann2019}.

Under the Rotating Wave Approximation (RWA) and when the two transmons are resonant, the single-photon hopping and cross-Kerr effects are observable with strengths $J_1$ and $V$, while the two-mode squeezing interaction is far off-resonant. In previous measurements on this device, the single-photon hopping and cross-Kerr interactions were shown to be highly tunable dependent on the choice of static coupler flux bias point with deep access to the regime $J_1>V$ \cite{Kounalakis2018}.

While the strengths of the linear and cross-Kerr couplings are ordinarily constrained by the engineered characteristics of the circuit and choice of flux bias point, by parametrically modulating the flux threading the SQUID loop of the coupler, one can access parameter regimes in which either the photon hopping or two-mode squeezing terms can be selectively activated. This selective activation enables one to induce---for even far off-resonant oscillators---linear interactions with strengths spanning a wide range of ratios $J_{1,2}/V$.

We consider the case in which the total magnetic flux threading the coupler contains a static DC component as well as a periodic AC component, where the total flux is given by
\begin{equation}
\begin{aligned}
    \Phi_C(t)=\Phi_{DC}+\Phi_{AC}\cos{(\omega_m t)}
    \label{phict}
\end{aligned}
\end{equation}
and $\omega_m$ is the frequency of the modulation. Provided that the strength of modulation is small relative to the bias point ($\sin{(\Phi_{DC})}\gg\sin{(\Phi_{AC})}$), we can insert Eq. \ref{phict} into Eq. \ref{ejc} and obtain a new expression for the Josephson energy of the coupler as
\begin{multline}
E_J^C(\Phi_C(t))\approx E_{J \rm max}^C|\cos{\bigg(\pi\frac{\Phi_{DC}}{\Phi_0}\bigg)}\\
-\pi\frac{\Phi_{AC}}{\Phi_0}\sin{\bigg(\pi\frac{\Phi_{DC}}{\Phi_0}\bigg)}\cos{(\omega_m t)}|\sqrt{1+d_c^2\tan^2{\bigg(\pi\frac{\Phi_{DC}}{\Phi_0}\bigg)}}\\
=E_{J,DC}^C+E_{J,AC}^C(t)
\label{cosine_exp}
\end{multline}
which is now comprised of a static term $E_{J,DC}^C$ and a time-dependent term $E_{J,AC}^C(t)$ due to the modulation.

After re-deriving the expressions for the hopping and two-mode squeezing interactions, it can be found that by modulating the coupler at the difference or sum frequency $\omega_m=|\omega_A\pm\omega_B|$, either interaction can be selectively activated for non-resonant oscillators as the coupling strengths under modulation are modified to 
\begin{multline}
J_1\rightarrow{}[J_{1,DC}+J_{AC}(e^{i\omega_m t}+e^{-i\omega_m t})]\\
\times(\hat{a}^{\dagger}\hat{b}e^{i(\omega_A-\omega_B) t}+\hat{a}\hat{b}^{\dagger}e^{-i(\omega_A-\omega_B)t})
\label{J1ac}
\end{multline}
\begin{multline}
J_2\rightarrow{}[J_{2,DC}+J_{AC}(e^{i\omega_m t}+e^{-i\omega_m t})]\\
\times(\hat{a}^{\dagger}\hat{b}^{\dagger}e^{i(\omega_A+\omega_B) t}+\hat{a}\hat{b}e^{-i(\omega_A+\omega_B) t})
\label{J2ac}
\end{multline}
where $J_{1,DC}$, $J_{2,DC}$ are as in Eq. \ref{J12} and the strength of the modulated interaction may be approximated as
\begin{equation}
\begin{aligned}
J_{AC}\approx\frac{\pi\Phi_{AC}}{4\sqrt{2}\Phi_0}\sin{\bigg(\pi\frac{\Phi_{DC}}{\Phi_0}\bigg)}E_{J \rm max}^C\bigg(\frac{E_{C}^AE_{C}^B}{\widetilde{E}_J^A\widetilde{E}_J^B}\bigg)^{1/4}
\label{Jac}
\end{aligned}.
\end{equation}
After applying the RWA, we may choose to activate either interaction with strength $J_{AC}$ depending on the frequency of modulation, while other terms not commensurate with the modulation become fast-rotating and play a negligible role in the system dynamics. The full form of Eq. \ref{Jac} and the contributions from higher-order interactions are shown in the supplemental material.

In order to measure the strength of the couplings under time-periodic pumping, we modulated the DC current supplied to the tunable coupler at a frequency $\omega_m$. The static component of the system under modulation can be written as
\begin{multline}
    \hat{H}_{DC}=\omega_A\hat{a}^\dagger \hat{a}+\frac{\alpha_A}{2}\hat{a}^\dagger \hat{a}^\dagger \hat{a} \hat{a}+\omega_B\hat{b}^\dagger \hat{b}+\frac{\alpha_B}{2}\hat{b}^\dagger \hat{b}^\dagger \hat{b} \hat{b} \\  
    +V\hat{a}^\dagger \hat{a} \hat{b}^\dagger \hat{b},
\end{multline}
with additional terms present depending on the frequency at which the coupler flux is modulated. When modulating at the red sideband (RSB), we have
\begin{equation}
\begin{aligned}
    \hat{H}_{RSB}=\hat{H}_{DC}+\hat{H}_{\Delta},
\label{HRSB}
\end{aligned}
\end{equation} 
\begin{equation}
\begin{aligned}
    \hat{H}_{\Delta}=J_{AC}(\hat{a}^\dagger \hat{b} + \hat{a}\hat{b}^\dagger)
\end{aligned}
\end{equation}
and when modulating at the blue sideband (BSB), we similarly obtain
\begin{equation}
\begin{aligned}
    \hat{H}_{BSB}=\hat{H}_{DC}+\hat{H}_{\Sigma},
\label{HBSB}
\end{aligned}
\end{equation}
\begin{equation}
\begin{aligned}
    \hat{H}_{\Sigma}=J_{AC}(\hat{a}^\dagger \hat{b}^\dagger + \hat{a}\hat{b})
\end{aligned}
\end{equation}
with additional contributions to the total interaction strengths from correlated hopping and squeezing terms as discussed in the supplemental material.

\begin{figure}[t]
    \captionsetup{justification = Justified,
              width=\linewidth}
    \includegraphics[width=\linewidth]{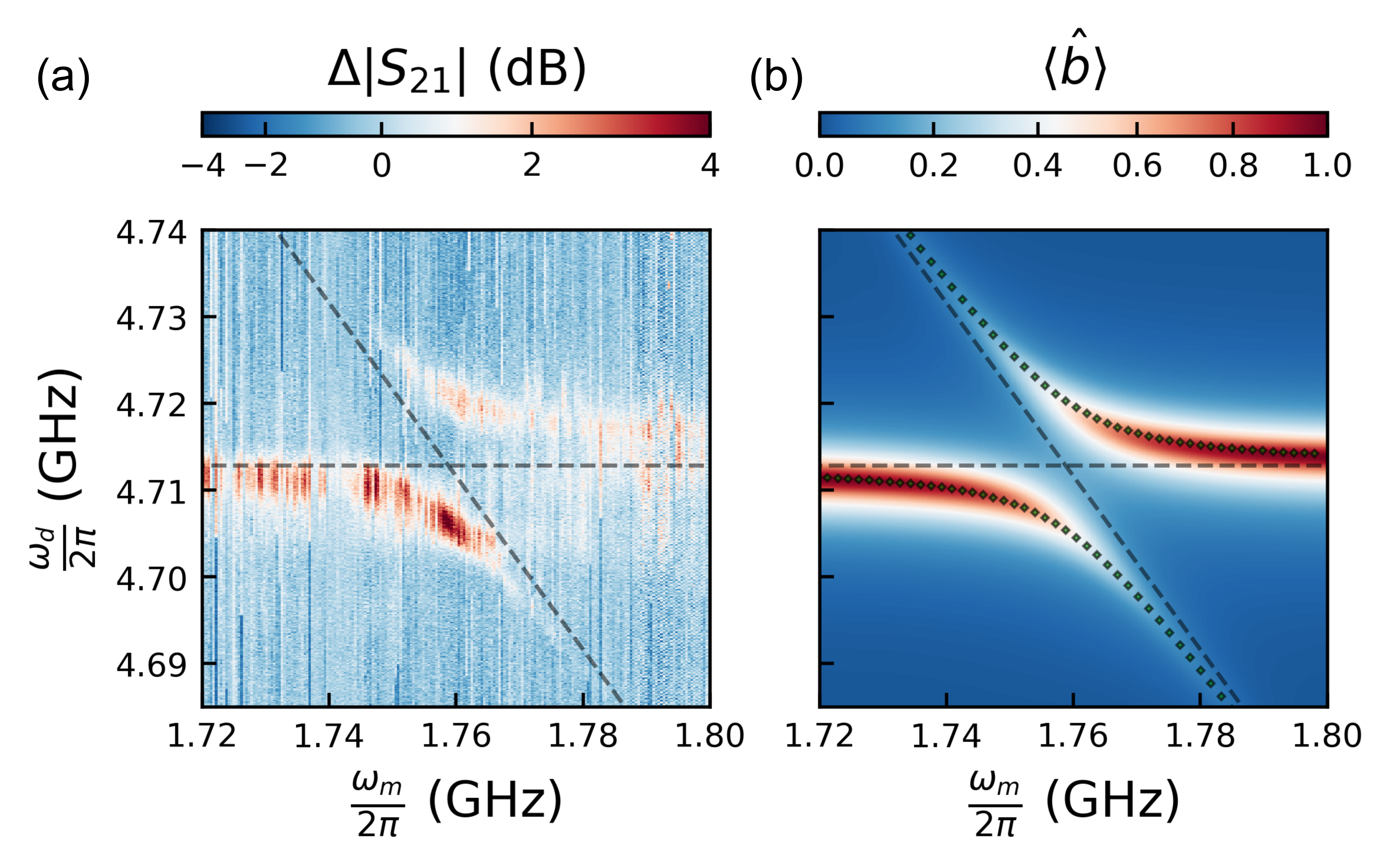}
    \caption{Single-photon hopping interaction induced by red sideband flux modulation of the coupler. (a) Change in transmission while driving transmon B and sweeping the modulation frequency of the DC signal incident on the tunable coupler through the red sideband of the two oscillators. The black dashed lines are guides for the eye. The horizontal dashed line is the first transition frequency of transmon B, and the diagonal dashed line is $(\omega_A-\omega_m)/2\pi$. (b) Eigenfrequencies obtained from fitting to the level repulsion model (markers) and the normalized expectation value of $\hat{b}$ obtained from a numerical quantum master equation simulation of the system.}
    \label{RSB Modulation}
\end{figure}

\begin{figure*}[t]
    \captionsetup{justification = Justified,
              width=\linewidth}
    \includegraphics[width=0.8\linewidth]{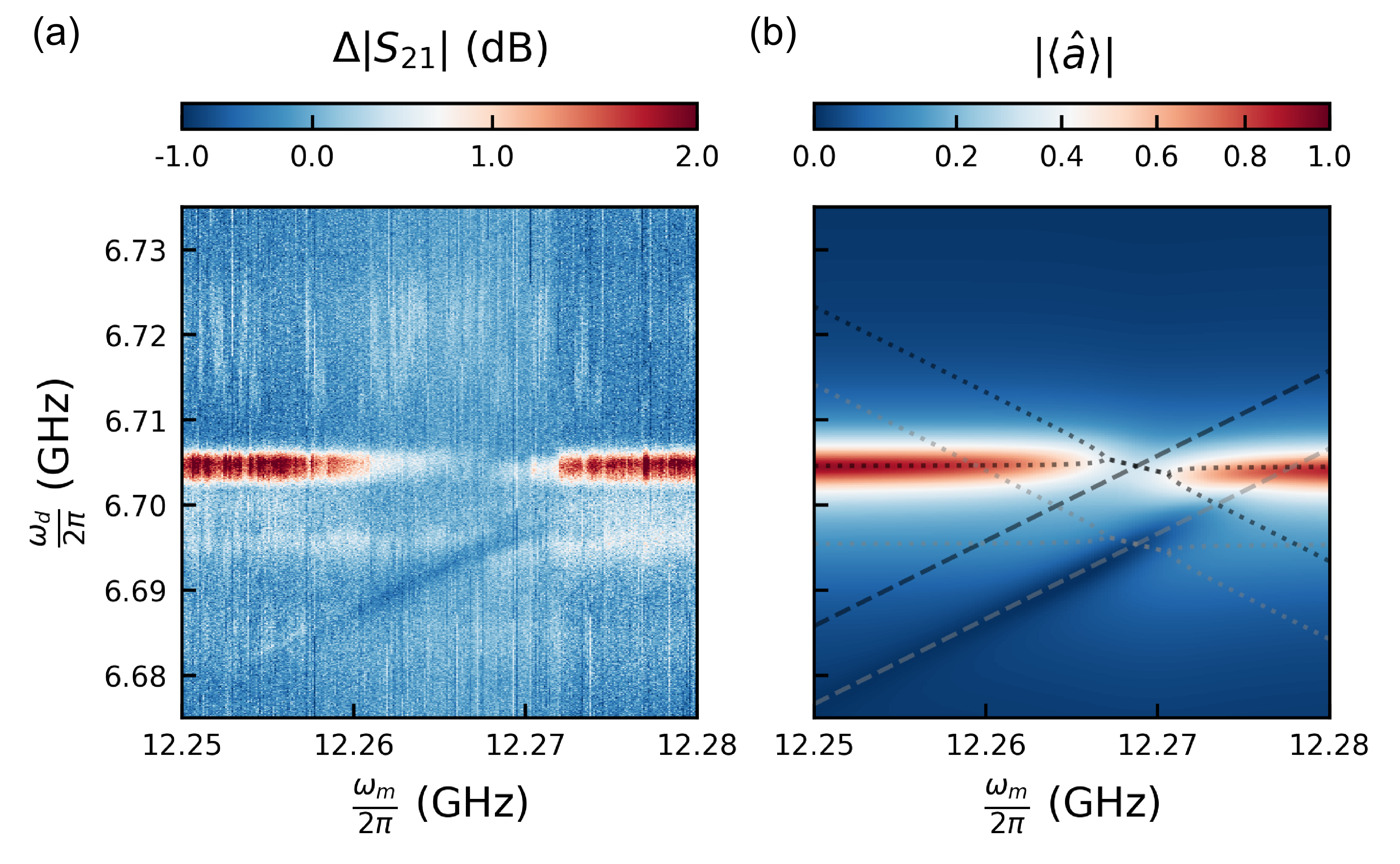}
    \caption{Two-mode squeezing interaction induced by blue sideband flux modulation of the coupler. (a) Change in transmission while driving transmon A and sweeping the modulation frequency of the DC bias incident on the tunable coupler through the blue sideband of the two oscillators. (b) The dashed lines are $(\omega_m-\omega_B)/2\pi$ and the same shifted vertically by $V/2\pi$. The dotted lines are the eigenfrequencies of the system determined from fits of the data to the analytical level attraction model. The underlying spectrum is the normalized expectation value of $\hat{a}$ obtained from a numerical simulation of a quantum master equation for the system.} 
    \label{BSB Modulation}
\end{figure*}

\section{Results}

In Fig.~\ref{RSB Modulation} we set $\omega_A/2\pi=6.472$ GHz, $\omega_B/2\pi=4.713$ GHz, $\Phi_{DC}=0.349\Phi_0$ and performed two-tone spectroscopy on transmon B while sweeping $\omega_m/2\pi$ through $(\omega_A-\omega_B)/2\pi$. As the modulation frequency approached the red sideband of the oscillators, we observed an avoided crossing from which we extracted a single-photon hopping interaction strength of $J_{AC}/2\pi=7.462$ MHz and a cross-Kerr strength of $V/2\pi=-6.543$ MHz from a fit of the data. Fit parameters were found given our observed oscillator frequencies and interaction strengths from an analytical level repulsion model and by comparison to numerical quantum master equation simulations of the system. The magnitude of the observed splitting reflects the strength of the exchange interaction between the two oscillators at the resonance condition met under parametric modulation.

Similarly, in Fig.~\ref{BSB Modulation} we set $\omega_A/2\pi=6.704$ GHz, $\omega_B/2\pi=5.573$ GHz and $\Phi_{DC}=0.215\Phi_0$ and performed two-tone spectroscopy on transmon A while sweeping $\omega_m/2\pi$ through the cross-Kerr shifted sum frequency $(\omega_A+\omega_B+V)/2\pi$. As the pump frequency crossed the blue sideband, we observed features associated with the phenomenon of level attraction occurring between the two oscillators. Again, from an analytical model and numerical simulations, we extracted a two-mode squeezing strength of $J_{AC}/2\pi=1.131$ MHz and a cross-Kerr strength of $V/2\pi=-9.158$ MHz with an additional cross-Kerr shifted transition visible below the frequency of the primary oscillator response. The cross-Kerr coupling yields both a small peak in transmission below the primary transition feature due to thermal population of the oscillator mode, as well as the shifted level attraction feature visible as a decrease in transmission.

In the level attraction region where frequency degeneracy of the eigenmodes is theoretically predicted, we observed the primary resonance feature disappear. In this same region, we observed the emergence of a dip in the transmission spectrum related to a loss of excited state population in transmon A. This absorption feature is shifted from the primary resonance by $V/2\pi$. It is associated with the microwave drive bringing the oscillator to its ground state from the excited state populated by the parametric modulation.

\begin{figure*}[t]
    \captionsetup{justification = Justified,
              width=\linewidth}
    \includegraphics[width=0.8\linewidth]{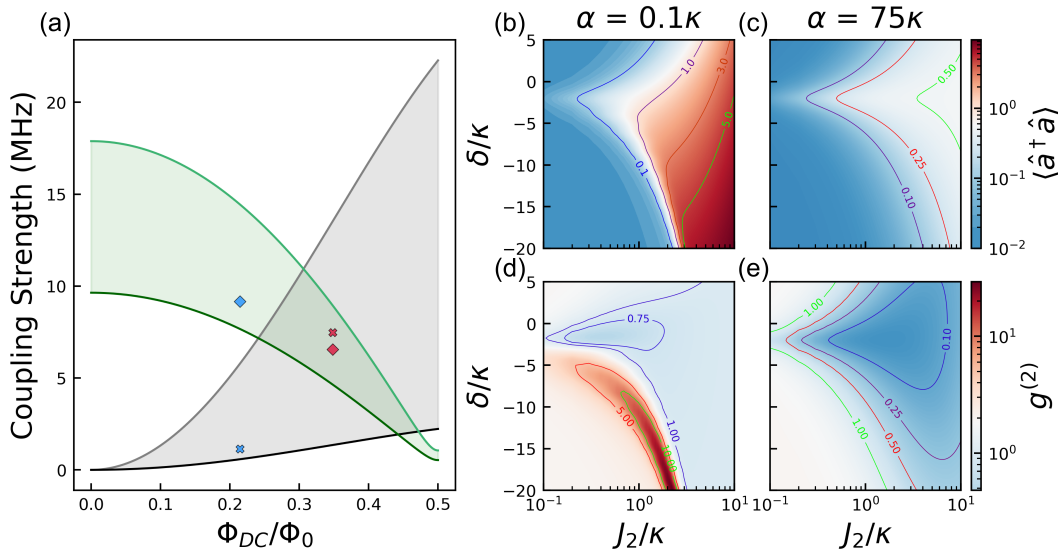}
    \caption{Interaction strengths and distinct response parameter regimes. (a) The calculated interaction strengths as the DC flux bias point of the coupler is changed. The gradient of curves in green shows the possible values of the cross-Kerr interaction for the range of $\Phi_{A,B}\in[0.0,0.5\Phi_0]$ at each value of $\Phi_{DC}$. The gradient of curves in gray shows the values of single-photon hopping or two-mode squeezing interactions for a range of modulation strengths between $\Phi_{AC} = \Phi_{DC}/100$ and $\Phi_{AC} = \Phi_{DC}/10$. The diamond ($V/2\pi$) and cross ($J_{AC}/2\pi$) markers indicate the extracted interaction strengths from Fig.~\ref{RSB Modulation} (red) and Fig.~\ref{BSB Modulation} (blue). (b), (c) The photon number expectation value for one mode of a system of two coupled Kerr-nonlinear oscillators subject to a drive-induced two-mode squeezing interaction where the strength of the interaction $J_2$, detuning of the drive from the sum frequency resonance condition $\delta$, and nonlinearity $\alpha$ are varied relative to the loss rates of the oscillators $\kappa$. (d), (e) The second-order correlation function of one of the oscillators as in (b), (c).} 
    \label{Modulation Strengths}
\end{figure*}

%Through parametric modulation, we observed interactions between two nonlinear oscillators in coupling regimes previously inaccessible with this device when changing the static flux bias point alone \cite{Kounalakis2018}.
When modulating the flux through the coupler, the strength of the single-photon hopping and two-mode squeezing interactions are to first order linearly dependent on the amplitude of the modulation signal and thus can be tuned to lower or higher interaction strengths relative to the cross-Kerr for a wide range of static biases. The dependence of the interaction strengths on bias point and modulation amplitude is shown in Fig.~\ref{Modulation Strengths}(a), where the green region indicates the range of theoretically achievable cross-Kerr values depending on the flux bias points of the transmons and coupler. In contrast, the grey region shows the viable values of $J_{AC}/2\pi$ for a range of modulation strengths.

While we demonstrated the ability to enter into this cross-Kerr dominant coupling regime, we also observed two-mode squeezing interactions, which are typically far off-resonant and fast-rotating in the frame of the oscillators. This entangling interaction generates coupled signal and idler modes and has been used to perform two-qubit gate (bSWAP) operations in the truncated qubit subspace \cite{Poletto2012, Roth2017, Bengtsson2018, Wustmann2019}. Activating this term enables the tuning of XX-YY interactions between the oscillators, broadening the array of systems such devices can effectively simulate. The modulated strength is tunable over a wide range, enabling the possibility for simulation of arbitrary XYZ spin-model Hamiltonians when coupled with the controllability demonstrated by the XX+YY and ZZ interactions \cite{Sameti2019}. 

Prospects for bichromatic flux pumps are also promising, where phase differences between simultaneously applied red and blue sideband pumps would allow for pure XX or YY interactions \cite{Sameti2019}. Such driven coupler schemes have been previously investigated in the context of Floquet engineering, in which tunable and selectively activatable interactions are integral to the proposed analog quantum simulation of Kitaev honeycomb models \cite{Sameti2019}. The nonstoquastic terms which a parametrically modulated tunable coupler can contribute to the system are also of great interest with respect to the study of possible quantum advantage over classical approaches in annealing protocols \cite{Jin2013, Hormozi2017, Ciani2021}.

These parametric interactions additionally provide insight into the dual phenomena of level repulsion and attraction. The red and blue sideband results illustrate the ability to transition between coherent (real) coupling and dissipative (imaginary) coupling simply with an applied modulation pump. Such couplings have long been the focus of study in magnonic, Bose-Einstein condensate, and optomechanical systems, typically operating in regimes where the nonlinearities of the oscillators are small \cite{Bernier2014, Xu2016, Zhang2017_2, Harder2018, Bernier2018, Bhoi2019, Yang2019, Yao2019, Wang2019_2, Yu2019}.

In the case of level attraction, a system of coupled linear oscillators exhibits a region of parametric instability with two exceptional points indicating the transition of the system to one with complex eigenfrequencies with opposite-sign imaginary components. In this situation, one eigenmode grows exponentially and becomes unstable while the other decays exponentially. Such a situation arises only when the dissipation rates of the two oscillator modes are commensurate and the coupling exceeds the oscillator decay rates \cite{Bernier2018}. For our two oscillators, the linewidths are each on the order of a few MHz for the flux points investigated, and the coupling strength can be tuned to less than or greater than the dissipation rates for typical transmon coherence times given the choice of modulation amplitude.

The same interaction was also previously observed between two coupled weakly nonlinear modes of a SQUID-terminated coplanar waveguide resonator subject to flux modulation \cite{Bengtsson2018}. Nondegenerate parametric oscillations were observed when the system was driven beyond the parametric instability threshold for a range of sufficient detunings and modulation strengths. In these systems, the self-Kerr and cross-Kerr couplings of the modes were often smaller than the decay rates, with large photon number states generable when driven above threshold \cite{Wustmann2013, Wustmann2017, Bengtsson2018, Wustmann2019}. Upon the application of an additional drive, such parametric oscillators can also become injection-locked to the drive \cite{Markovic2019}.

The main distinctions between the behavior outlined above and what we observe in Fig.~\ref{BSB Modulation} are that the oscillators measured are strongly nonlinear and that we do not observe a region of parametric instability due to the low strength of the two-mode squeezing interaction relative to the self-Kerr-nonlinearities of the oscillators. Further, our oscillators are additionally cross-Kerr coupled which yields a frequency shift of the spectroscopic features. In order to investigate these distinctions and better understand the contributions of the self-Kerr and cross-Kerr terms to the phenomenon of level attraction between strongly nonlinear oscillators, we numerically simulated the system for a variety of parameters. 

In Fig.~\ref{Modulation Strengths}(b-e), we show for one mode the photon number expectation values $\langle a^\dagger a\rangle$ and second-order correlation function $g^{(2)}$ in a system of two Kerr-nonlinear oscillators as in Eq.~\ref{HBSB} where the correlated squeezing terms are set to zero, determined from quantum master equation simulations \cite{QuTiPCite}. We set $V=-2\kappa$ and vary the strength of the two-mode squeezing term $J_2$ and detuning of the modulation frequency from the sum frequency resonance condition $\delta$ for the case of weakly nonlinear oscillators $\alpha=0.1\kappa$ and strongly nonlinear oscillators $\alpha=75\kappa$. For the weakly nonlinear system, as the strength of the two-mode squeezing interaction increases, the parametric response region, which provides an increased photon number, shifts to large, negative detunings. Additionally, a sudden peak in $g^{(2)}$ bounds the parametric response region from below, which is a known marker of a phase transition in KNOs~\cite{Bartolo2016}.

In contrast, for strongly nonlinear oscillators, such as transmon qubits, for $J_2<\alpha$ and in a region centered about the cross-Kerr shifted sum frequency resonance condition, the photon number expectation and second-order correlation function remain below one. In this case, the two-mode squeezing interaction acts effectively on the qubit subspace alone, generating an XX-YY interaction. The large self-Kerr-nonlinearities of the oscillators prevents the system from reaching a parametric instability as in the case of the linear and weakly nonlinear two-mode squeezed systems, instead generating a low photon number entangled state.

%, except for sufficiently large values of $J_2$ and $\delta$ relative to the nonlinearity $E_C$.

%In the supplemental material, we display the relative change in the occupation of the first two oscillator energy levels with the self-Kerr and cross-Kerr terms included in the system Hamiltonian. We find that the negative cross-Kerr-nonlinearity shifts down the energy of the $\ket{11}$ state, leading to the appearance of a ``ghost" level attraction window corresponding to the population being driven out of the jointly excited state, which the parametric modulation had populated.

\section{Summary - Outlook}

In summary, we have demonstrated the operation of a transmon-based circuit containing a flux-tunable coupler enabling access into different coupling regimes, including cross-Kerr couplings ($V$), between two nonlinear oscillators. By parametrically modulating the inductance of the coupler SQUID loop with an applied time-dependent magnetic field, we can selectively activate either a single-photon hopping coupling ($J_1$) or two-mode squeezing coupling ($J_2$) between two transmon qubits.
The coupling strengths can be tuned by choosing different DC flux bias points and modulation amplitude. In combination with previously reported measurements of strong resonant single-photon hopping interactions on this device~\cite{Kounalakis2018}, this scheme gives access to parameter regimes where $J_{1,2}> V$, $J_{1,2}\approx V$ and $J_{1,2}< V$.

This tunability allows for the simulation of various systems, including Ising ZZ, Bose-Hubbard, and Heisenberg XXZ models~\cite{Jin2013_2, Georgescu2014, Hartmann2016, Sameti2019}. The ability to tune into and out of these regimes is of particular interest for analog quantum simulations, where such superconducting devices can be made to emulate a variety of physical systems with solely in-situ control and a broad range of coupling strengths achievable. Our circuit model predicts that further measurements using modulated couplers operating in different conditions may be used to activate more regimes, such as photon-pair tunnelling, correlated photon hopping, and photon-pressure interactions. Using asymmetric nonlinear elements such as SNAILs would also enable the simulation of more exotic interactions and simultaneously enable tuning of several device parameters, such as self-Kerr terms, which can be tuned from negative to positive values~\cite{Frattini2017, Lu2023}. 
The broad selectivity of system parameters in tunably coupled nonlinear oscillators is of particular interest due to the ability to investigate instability regimes, applications to parametric amplification, driven-dissipative interactions, as well as exploring non-Hermitian Hamiltonians~\cite{Metelmann2014, Hartmann2016, Hurst2022}. Exquisite control over these interactions would enable direct investigation of coherent and dissipative couplings between nonlinear oscillators and bring predicted applications in topological energy transfer, quantum sensing, and nonreciprocal photon transmission closer to experimental realization~\cite{Kohler2018, Harder2018, Bernier2018_2, Bhoi2019, Yao2019}.

Finally, under tunable blue sideband modulation, we have observed level attraction between the two nonlinear oscillators. Interestingly, the behavior of the system differs from previously established theoretical descriptions and experimental observations of linear systems exhibiting level attraction \cite{Bernier2018, Wang2019_2, Yu2019}. Using an extension of existing methods and numerical simulations, we were able to determine that the cross-Kerr coupling yields an additional shifted spectroscopic feature of level attraction and that signatures of level attraction can be observed in the absence of parametric instability in the case of strongly Kerr-nonlinear oscillators.
The dual phenomena of level repulsion and level attraction have been previously investigated in a broad array of platforms ranging from Bose-Einstein condensates to magnonic and optomechanical systems operating in various parameter regimes characterized by the resonance frequencies, coupling strengths and decay rates of the constituent oscillators~\cite{Bernier2014, Eichler2014, Xu2016, Zhang2017_2, Bernier2018, Harder2018, Grigoryan2018, Bhoi2019, Yu2019, Yang2019, Miri2019, Wang2019_2, Wang2020}. Furthermore, dissipative couplings and two-mode squeezing interactions giving rise to level attraction are particularly useful for enabling quantum-limited nondegenerate parametric amplification as well as performing two-qubit gate operations~\cite{Eichler2014, Roth2017, Bengtsson2018}.

%Under bichromatic $2\omega_A$ and $2\omega_B$ pumps, such tunably coupled oscillators would also be useful for investigating stabilized dissipation in driven Bose-Hubbard systems \cite{Mamaev2018}. Further, the independently tunable strengths of the photon hopping and two-mode squeezing interactions should allow for investigation into novel parametric interaction regimes such as those studied in optomechanical systems \cite{Aldana2013, Metelmann2014}.

\subsection*{Contributions}\label{contribs4}
J.D.K., G.B., and M.K. carried out the theoretical analysis. M.K. designed and fabricated the device in the group of Leo DiCarlo. J.D.K., F.F.S, and M.K. conducted the measurements. J.D.K and G.B. performed the simulations. M.K. and G.A.S. conceived the experiment. C.A.P., M.K., and G.A.S. supervised the project. J.D.K. wrote the manuscript with input from all authors.

\begin{acknowledgments}
J.D.K., M.K., and G.A.S. acknowledge financial support by the EU program H2020-FETOPEN project 828826 Quromorphic. M.K. acknowledges financial support from the Netherlands Organisation for Scientific Research (NWO/OCW). C.A.P. acknowledges the support of the Natural Sciences and Engineering Research Council of Canada (NSERC) (PDF-567689-2022) and the Novo Nordisk Foundation, NNF Quantum Computing Programme.
\end{acknowledgments}

% \putbib[apssamp]
\bibliography{apssamp_main}
 
%%%%%%%%%% Merge with supplemental materials %%%%%%%%%%
\clearpage
\widetext
\begin{center}
\textbf{\large Supplementary Information: Flux-modulated tunable interaction regimes in two strongly nonlinear oscillators}

\vspace{11pt}

J.~D. Koenig,$^{1}$ G. Barbieri,$^1$ F. Fani Sani,$^1$ C.~A. Potts,$^{1,}$$^{2,}$$^3$ M. Kounalakis,$^{1,}$$^4$ and G.~A. Steele$^{1}$

\vspace{11pt}

\footnotesize
$^1$ \textit{Kavli Institute of Nanoscience, Delft University of Technology, PO Box 5046, 2600 GA Delft, The Netherlands}

$^2$ \textit{Niels Bohr Institute, University of Copenhagen, Blegdamsvej 17, 2100 Copenhagen, Denmark}

$^3$ \textit{NNF Quantum Computing Programme, Niels Bohr Institute, University of Copenhagen, Denmark}

$^4$ \textit{Luxembourg Institute of Science and Technology (LIST), 4362, Esch-sur-Alzette, Luxembourg}

% $^*$ \textit{j.d.koenig@tudelft.nl}

% $^\dagger$ \textit{g.a.steele@tudelft.nl}

\end{center}

%%%%%%%%%% Merge with supplemental materials %%%%%%%%%%
%%%%%%%%%% Prefix a "S" to all equations, figures, tables and reset the counter %%%%%%%%%%
\setcounter{equation}{0}
\setcounter{figure}{0}
\setcounter{table}{0}
\makeatletter
\renewcommand{\theequation}{S\arabic{equation}}
\renewcommand{\thefigure}{S\arabic{figure}}
\renewcommand{\bibnumfmt}[1]{[S#1]}
% \renewcommand{\citenumfont}[1]{S#1}
%%%%%%%%%% Prefix a "S" to all equations, figures, tables and reset the counter %%%%%%%%%%

\section{Experimental Setup}\label{SI-4-Setup}
\begin{figure}[b!]
    %\begin{subfigure}
    \includegraphics[width=\linewidth]{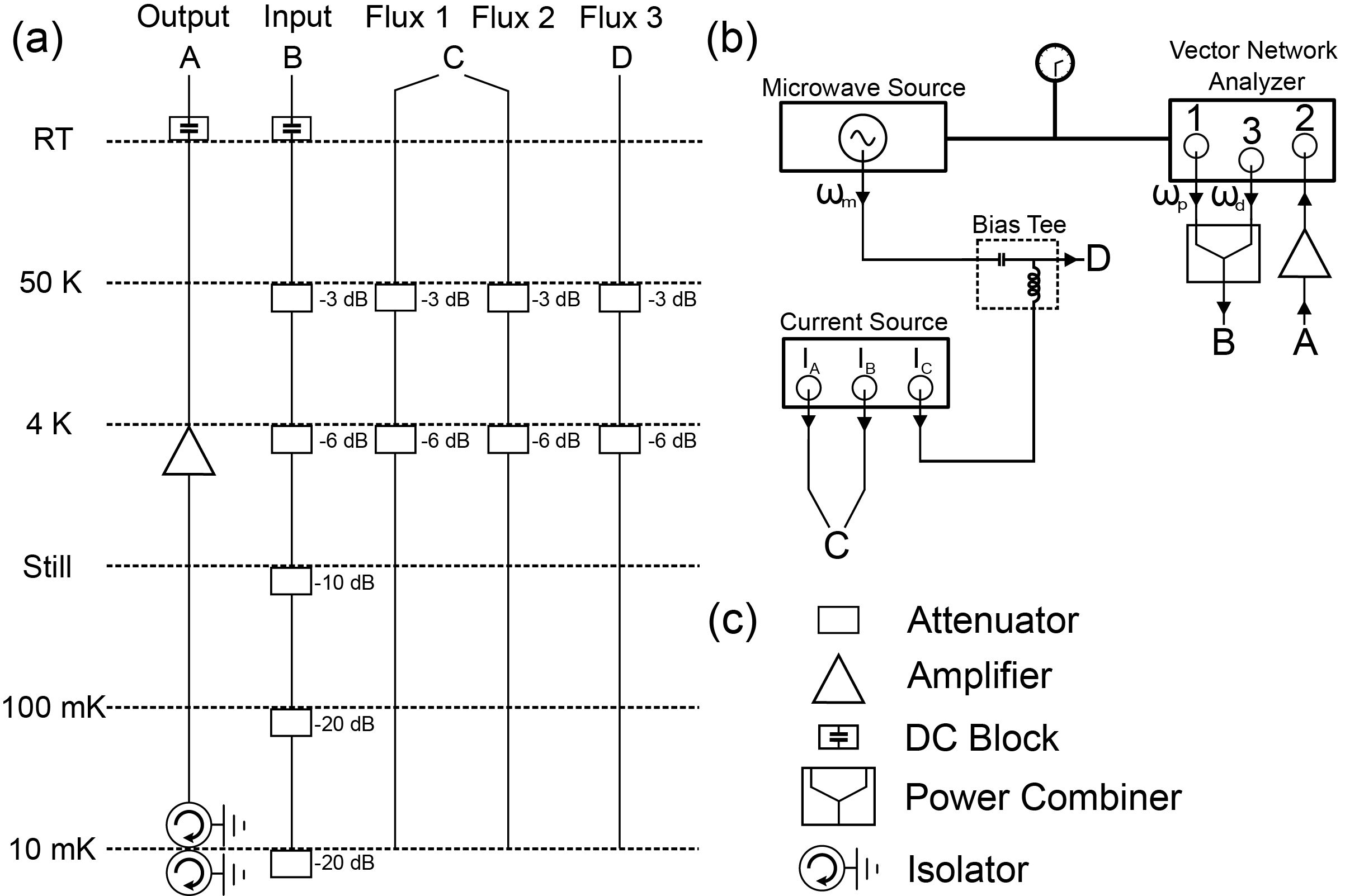}
    \caption{Measurement setup for the experiment. (a) Wiring configuration from the top of the dilution refrigerator to the device. The input and flux lines are attenuated before reaching the device, while the outgoing signal passes through two isolators and is amplified before returning to the network analyzer. (b) Room temperature wiring configuration. The device probe and drive signals are sent from ports 1 and 3 of the vector network analyzer (VNA), while the returning signal is further amplified before arriving at port 2. The DC signals sent to the flux lines are produced by a current source, with the coupler current modulated by a signal from an additional microwave source. The microwave source and VNA share a common reference clock signal. (c) Legend for microwave components.}
    \label{Wiring}
\end{figure}

The device in Fig.~\ref{Circuit Diagram}(b) of the main text contains a transmission line with input and output for probing and driving of the qubits, two coplanar waveguide readout resonators, two transmon qubits with dedicated flux lines and (unused) drive lines, and the tunable coupler with a dedicated flux line. The device is the same as in Ref. \cite{Kounalakis2018}, with elements defined on a NbTiN film deposited on a Si substrate, with the chip wirebonded to a printed circuit board mounted inside of a copper box, and the entire unit housed in a mu-metal shield for protection against external magnetic fields.

The measurement setup shown in Fig.~\ref{Wiring} consists of a Keysight PNA N5222A network analyzer connected to the device transmission line for spectroscopy measurements as well as a Keysight E8257D signal generator connected to a bias tee to supply the modulation signal to the coupler flux line. Double DC blocks were installed on the input and output lines at room temperature. The DC currents were supplied by a QuTech SPI Rack S4g current source module to the three flux lines. For amplification of the signal returning from the circuit, we used a Low Noise Factory cryogenic amplifier LNF-LNC4\verb|_|8C and a room temperature Narda-MITEQ amplifier in the 4-8 GHz range.

\section{Circuit Quantization}\label{SI-4-Circuit}
Following Ref.~\cite{Kounalakis2018}, we begin by treating the circuit in the harmonic limit by neglecting the nonlinear contributions of the inductors. We can construct a Lagrangian for our circuit by defining the node basis as $\mathbf{\Phi}^T$=$[\Phi_1,\Phi_2,\Phi_3,\Phi_4]$ with
\begin{equation}
\begin{aligned}
\mathcal{L}=\mathcal{E_C}-\mathcal{E_L}=\frac{1}{2}\dot{\mathbf{\Phi}}^T[\mathbf{C}]\dot{\mathbf{\Phi}}-\frac{1}{2}\mathbf{\Phi}^T[\mathbf{L}^{-1}]\mathbf{\Phi}
\end{aligned}
\end{equation}
where the capacitance and inductance matrices are written as
\begin{equation} 
\begin{aligned}
[\mathbf{C}]=
\begin{pmatrix}
C+C_{1g} & -C & 0 & 0\\
-C & C+C_{2g}+C_c & -C_c & 0\\
0 & -C_c & C+C_{2g}+C_c & -C\\
0 & 0 & -C & C+C_{1g}
\end{pmatrix}
\end{aligned} 
\end{equation}
\begin{equation} 
\begin{aligned}
[\mathbf{L}^{-1}]=
\begin{pmatrix}
1/L_1 & -1/L_1 & 0 & 0\\
-1/L_1 & 1/L_1+1/L_c & -1/L_c & 0\\
0 & -1/L_c & 1/L_2+1/L_c & -1/L_2\\
0 & 0 & -1/L_2 & 1/L_2
\end{pmatrix}
\end{aligned}.
\end{equation}

We perform a change of basis to express the first two normal modes of the circuit as what will become the transmon modes, $\Psi_A\equiv\Phi_1-\Phi_2$ and $\Psi_B\equiv\Phi_3-\Phi_4$. There exists a third normal mode associated with the coupler, as charge oscillations ``slosh" across the circuit. We define this mode as $\Psi_S\equiv\frac{1}{2}(\Phi_1+\Phi_2-\Phi_3-\Phi_4)$. There also exists a final zero-frequency ``rigid" mode associated with the charging of all capacitors in unison, defined as $\Psi_R\equiv\frac{1}{2}(\Phi_1+\Phi_2+\Phi_3+\Phi_4)$. The change of basis from $\mathbf{\Phi}^T$ to $\mathbf{\Psi}^T$=$[\Psi_A,\Psi_B,\Psi_S,\Psi_R]$ leads to the redefinition of the capacitance matrix as 
\begin{equation} 
\begin{aligned}
[\textbf{C}^\prime]=
\begin{pmatrix}
\frac{C_{1g}+C_{2g}}{8} & 0 & \frac{C_{1g}-C_{2g}}{8} & \frac{C_{1g}-C_{2g}}{8}\\
0 & \frac{C_{1g}+C_{2g}+2C_c}{8} & \frac{-C_{1g}+C_{2g}+2C_c}{8} & \frac{C_{1g}-C_{2g}-2C_c}{8}\\
\frac{C_{1g}-C_{2g}}{8} & \frac{-C_{1g}+C_{2g}+2C_c}{8} & C+\frac{C_{1g}+C_{2g}+2C_c}{4} & -\frac{C_c}{4}\\
\frac{C_{1g}-C_{2g}}{8} & \frac{C_{1g}-C_{2g}-2C_c}{8} & -\frac{C_c}{4} &  C+\frac{C_{1g}+C_{2g}+2C_c}{4} 
\end{pmatrix}
\end{aligned}. 
\end{equation}

We now treat the inductors as nonlinear elements with energy in reduced units of flux quanta, $\psi_i=\frac{2\pi}{\Phi_0}\Psi_i$. After defining the conjugate momenta as $Q_i=\frac{\partial\mathcal{L}}{\partial\psi_i}$, we perform a Legendre transformation and obtain 
\begin{multline}
H=\frac{Q_A^2}{2\tilde{C}}+\frac{Q_B^2}{2\tilde{C}}+\frac{Q_S^2}{2\tilde{C}_S}+\frac{Q_R^2}{2\tilde{C}_R}+\frac{C_cC_{1g}^2}{4Det[\mathbf{C}^\prime]}Q_AQ_B+\frac{1}{\tilde{C}_{ABS}}Q_S(Q_A-Q_B)\\
+\frac{1}{\tilde{C}_{ABR}}Q_R(Q_A+Q_B)-E_J^A\cos(\psi_A)-E_J^B\cos(\psi_B)-E_J^C\cos\bigg(\frac{\psi_A-\psi_B}{2}-\psi_S\bigg)
\end{multline}
where
\begin{equation} 
\begin{aligned}
\tilde{C}=4Det[\mathbf{C}^\prime] [C_{1g}C_{2g}(C_{1g}+C_{2g})+C_{1g}C_c(C_{1g}+2C_{2g})+C(C_{1g}+C_{2g})(C_{1g}+C_{2g}+2C_c)]^{-1}
\end{aligned} 
\end{equation}
\begin{equation} 
\begin{aligned}
\tilde{C}_S=2\frac{C_{1g}(C_{2g}+2C_c)+C(C_{1g}+C_{2g}+2C_c)}{4C+C_{1g}+C_{2g}+2C_c}
\end{aligned} 
\end{equation}
\begin{equation} 
\begin{aligned}
\tilde{C}_R=2\frac{C_{1g}C_{2g}+C(C_{1g}+C_{2g})}{4C+C_{1g}+C_{2g}}
\end{aligned} 
\end{equation}
\begin{equation} 
\begin{aligned}
\tilde{C}_{ABS}=2\frac{C_{1g}(C_{2g}+2C_c)+C(C_{1g}+C_{2g}+2C_c)}{C_{2g}-C_{1g}+2C_c}
\end{aligned} 
\end{equation}
\begin{equation} 
\begin{aligned}
\tilde{C}_{ABR}=2\frac{C_{1g}C_{2g}+C(C_{1g}+C_{2g})}{C_{2g}-C_{1g}}
\end{aligned} 
\end{equation}
\begin{equation} 
\begin{aligned}
Det[\mathbf{C}^\prime]=\frac{C_{1g}C_{2g}+C(C_{1g}+C_{2g})}{4(C_{1g}(C_{2g}+2C_c)+C(C_{1g}+C_{2g}+2C_c))}
\end{aligned}.
\end{equation}

We can neglect the rigid mode entirely by shifting the charging energies of the transmons and coupler, account for the addition of inductive energy from the coupler to each transmon by taking $E_J^i\rightarrow E_J^i+E_J^C/4$ from here on, and express the Hamiltonian in the number basis as $N=\frac{1}{2e}Q$. We expand the cosine terms above and retain terms to fourth order, obtaining $H=H_T+H_S$ where
\begin{multline}
H_T=4E_CN_A^2+\frac{E_J^A}{2}\psi_A^2-U_A\psi_A^4+4E_CN_B^2+\frac{E_J^B}{2}\psi_B^2-U_B\psi_B^4+E_C^CN_AN_B\\
+\frac{4e^2}{\tilde{C}_{ABS}}N_S(N_A-N_B)-\frac{E_J^C}{4}\psi_A\psi_B-\frac{E_J^C}{2}(\psi_A-\psi_B)\psi_S\\
-\frac{E_J^C}{64}\psi_A^2\psi_B^2+\frac{E_J^C}{96}(\psi_A^3\psi_B+\psi_A\psi_B^3)-\frac{E_J^C}{16}(\psi_A-\psi_B)^2\psi_S^2+\frac{E_J^C}{12}(\psi_A-\psi_B)\psi_S^3
\end{multline}
\label{Ht}
\begin{equation} \begin{aligned}
H_S=4E_C^SN_S^2+\frac{E_J^C}{2}\psi_S^2-\frac{E_J^C}{24}\psi_S^4
\end{aligned} 
\end{equation}
\label{Hs}
where $H_T$ is the transmon Hamiltonian, $H_S$ is the ``sloshing" mode Hamiltonian, e is the electron charge, $E_C=\frac{e^2}{2}(\frac{1}{\tilde{C}}-\frac{\tilde{C}_R}{\tilde{C}_{ABR}^2})$, $E_C^C=e^2(\frac{C_cC_{1g}^2}{4Det\textbf{[$C^\prime$]}}-\frac{\tilde{C}_R}{\tilde{C}_{ABR}^2})$, $E_C^S=\frac{e^2}{2\tilde{C}_S}$, and $U_i=E_J^i/24+E_J^C/384$. 

We can now move to the harmonic oscillator basis by defining 
\begin{equation} 
\begin{aligned}
\psi_A=\bigg(2\frac{E_C}{E_J^A}\bigg)^{1/4}(\hat{a}^\dagger+\hat{a})
\end{aligned} 
\end{equation}
\begin{equation} 
\begin{aligned}
\psi_B=\bigg(2\frac{E_C}{E_J^B}\bigg)^{1/4}(\hat{b}^\dagger+\hat{b})
\end{aligned} 
\end{equation}
\begin{equation} 
\begin{aligned}
\psi_S=\bigg(2\frac{E_C^S}{E_J^C}\bigg)^{1/4}(\hat{s}^\dagger+\hat{s})
\end{aligned} 
\end{equation}
\begin{equation} 
\begin{aligned}
N_A=i\bigg(\frac{E_J^A}{32E_C}\bigg)^{1/4}(\hat{a}^\dagger-\hat{a})
\end{aligned} 
\end{equation}
\begin{equation} 
\begin{aligned}
N_B=i\bigg(\frac{E_J^B}{32E_C}\bigg)^{1/4}(\hat{b}^\dagger-\hat{b})
\end{aligned} 
\end{equation}
\begin{equation} 
\begin{aligned}
N_S=i\bigg(\frac{E_J^C}{32E_C^S}\bigg)^{1/4}(\hat{s}^\dagger-\hat{s})
\end{aligned} 
\end{equation}
for the two transmons A and B, and the sloshing mode given by S. The terms proportional to $N^2, \psi^2,$ and $\psi^4$ describe uncoupled Duffing oscillators. The interaction terms solely between the transmons may be expressed in this basis as
\begin{equation} 
\begin{aligned}
E_C^CN_AN_B=\bigg(\frac{E_J^AE_J^B}{32E_C^2}\bigg)^{1/4}[(a^\dagger b + ab^\dagger)- (a^\dagger b^\dagger + ab)]
\end{aligned} 
\end{equation}
\begin{multline}
\frac{E_J^C}{64}\psi_A^2\psi_B^2=\frac{E_J^CE_C}{8\sqrt{E_J^AE_J^B}}[a^\dagger ab^\dagger b + \frac{1}{2}(a^\dagger a + b^\dagger b) +  \frac{1}{4}(a^{\dagger 2} b^2+a^2 b^{\dagger 2})]\\
+ \frac{1}{4}(a^{\dagger 2} b^{\dagger 2} + a^2 b^2)+\frac{1}{4}a^\dagger a(b^{\dagger 2}+b^2)+\frac{1}{4}b^\dagger b(a^{\dagger 2}+a^2)]
\end{multline}
\begin{multline}
\frac{E_J^C}{96}\psi_A^3\psi_B=\frac{E_J^CE_C}{48((E_J^A)^3E_J^B)^{1/4}}[(a^\dagger b + a b^\dagger) + (a^\dagger b^\dagger + ab)+2(a^\dagger + a)a^\dagger a(b^\dagger+b)\\
+(a^{\dagger 3}b+a^3b^\dagger)+(a^{\dagger 3}b^\dagger+a^3b)]
\end{multline}
\begin{multline}
\frac{E_J^C}{96}\psi_A\psi_B^3=\frac{E_J^CE_C}{48(E_J^A(E_J^B)^3)^{1/4}}[(a^\dagger b + a b^\dagger) + (a^\dagger b^\dagger + ab)+2(a^\dagger + a)b^\dagger b(b^\dagger+b)\\
+(a^\dagger b^3+ab^{\dagger 3})+(a^{\dagger}b^{\dagger 3}+ab^3)]
\end{multline}
while those also involving the sloshing mode are written as
\begin{multline}
    N_S(N_A-N_B)=\frac{4e^2}{\tilde{C}_{ABS}}[\bigg(\frac{E_J^CE_J^A}{32E_C^SE_C}\bigg)^{1/4}((a^\dagger s + as^\dagger)-(a^\dagger s^\dagger+as))\\
    -\bigg(\frac{E_J^CE_J^B}{32E_C^SE_C}\bigg)^{1/4}((b^\dagger s + bs^\dagger)-(b^\dagger s^\dagger+bs))]
\end{multline}
\begin{equation} 
\begin{aligned}
    \frac{E_J^C}{12}(\psi_A-\psi_B)\psi_S^3=\frac{1}{6}(E_J^CE_C(E_C^S)^3)^{1/4}[\frac{1}{(E_J^A)^{1/4}}(a^{\dagger}s^{3}+as^{\dagger 3})+\frac{1}{(E_J^B)^{1/4}}(b^\dagger s^3 s+bs^{\dagger 3})]
\end{aligned}
\end{equation}
\begin{equation} 
\begin{aligned}
    \frac{E_J^C}{16}\psi_A^2\psi_S^2=\frac{1}{4}\sqrt{\frac{E_J^CE_CE_C^S}{E_J^A}}(a^\dagger a + s^\dagger s +2a^\dagger a s^\dagger s)
\end{aligned} 
\end{equation}
\begin{equation} 
\begin{aligned}
    \frac{E_J^C}{16}\psi_B^2\psi_S^2=\frac{1}{4}\sqrt{\frac{E_J^CE_CE_C^S}{E_J^B}}(b^\dagger b + s^\dagger s +2b^\dagger b s^\dagger s)
\end{aligned}
\end{equation}
\begin{equation} 
\begin{aligned}
    \frac{E_J^C}{8}\psi_A\psi_B\psi_S^2=\frac{1}{2}\frac{\sqrt{E_J^CE_CE_C^S}}{(E_J^AE_J^B)^{1/4}}(s^\dagger s+\frac{1}{2})[(a^\dagger b +ab^\dagger)+(a^\dagger b^\dagger+ab)]
\end{aligned}. 
\end{equation}

Notably, the sloshing mode contributes small corrections to the transmon frequencies and hopping/squeezing interactions, as well as cross-Kerr effects between the transmons and the sloshing mode. Most transitions of the sloshing mode are far off-resonant from the transmon transition frequencies. However, the 0-3 transition of the sloshing mode is near-resonant with the range of qubit frequencies for $\Phi_{DC}\approx0.3\Phi_0$ \cite{Kounalakis2018}.

In the context of Eq. \ref{HRSB} and Eq. \ref{HBSB} from the main text, we then have for \textit{i}$\in\{A,B\}$:
\begin{equation} 
\begin{aligned}
    \omega_i=\sqrt{8E_J^iE_C}+\alpha_i+\frac{1}{2}(V+V_{iS})
\end{aligned} 
\end{equation}
\begin{equation} 
\begin{aligned}
    \alpha_i=-E_C[1-\frac{E_J^C}{16}\bigg(\frac{1}{E_J^i}-\frac{1}{\sqrt{E_J^A E_J^B}}\bigg)]
\end{aligned} 
\end{equation}
\begin{equation} 
\begin{aligned}
    \omega_S=\sqrt{8E_J^CE_C^S}+\alpha_S-\frac{1}{4}\sqrt{E_J^CE_CE_C^S}\bigg(\frac{1}{\sqrt{E_J^A}}+\frac{1}{\sqrt{E_J^B}}\bigg)
\end{aligned} 
\end{equation}
\begin{equation} 
\begin{aligned}
    \alpha_S=-E_C^S
\end{aligned} 
\end{equation}
\begin{equation} 
\begin{aligned}
    V=-\frac{E_J^CE_C}{8\sqrt{E_J^AE_J^B}}
\end{aligned} 
\end{equation}
\begin{multline}
    J_{AC}=\frac{\pi\Phi_{AC}}{8\Phi_0}E_{J \rm max}^C\sin\bigg(\frac{\pi\Phi_{DC}}{\Phi_0}\bigg)\bigg[\bigg(\frac{4E_C^2}{E_J^AE_J^B}\bigg)^{1/4}-\frac{E_C}{12}\bigg(\frac{1}{((E_J^A)^3E_J^B)^{1/4}}+\frac{1}{(E_J^A(E_J^B)^3)^{1/4}}\bigg)\\
    -\bigg(\frac{E_C^2(E_C^S)^2}{(E_J^C)^2E_J^AE_J^B}\bigg)^{1/4}\bigg]
\label{fullJac}
\end{multline} 
\begin{equation} 
\begin{aligned}
    J_{n_A}=\frac{\pi\Phi_{AC}}{24\Phi_0}E_{J \rm max}^C\sin{\bigg(\frac{\pi\Phi_{DC}}{\Phi_0}\bigg)}E_C\bigg(\frac{1}{(E_J^A)^3E_J^B}\bigg)^{1/4}
\end{aligned} 
\end{equation}
\begin{equation} 
\begin{aligned}
    J_{n_B}=\frac{\pi\Phi_{AC}}{24\Phi_0}E_{J \rm max}^C\sin{\bigg(\frac{\pi\Phi_{DC}}{\Phi_0}\bigg)}E_C\bigg(\frac{1}{E_J^A(E_J^B)^3}\bigg)^{1/4}
\end{aligned} 
\end{equation}
\begin{equation} 
\begin{aligned}
    J_{n_S}=\frac{\pi\Phi_{AC}}{2\Phi_0}E_{J \rm max}^C\sin{\bigg(\frac{\pi\Phi_{DC}}{\Phi_0}\bigg)}\bigg(\frac{E_C^2(E_C^S)^2}{(E_J^C)^2E_J^AE_J^B}\bigg)^{1/4}
\end{aligned} 
\end{equation}
\begin{equation} 
\begin{aligned}
    V_{iS}=-\frac{1}{2}\sqrt{E_CE_C^S\frac{E_J^C}{E_J^i}}
\end{aligned} 
\end{equation}
where all $E_J^i$ for the transmons and coupler are dependent on their static flux biases. For simulations of the system, we use $E_J^A/h=23.01$ GHz, $E_J^B/h=23.01$ GHz, $E_J^C/h=7.75$ GHz, $d_{A,B}\approx0.50$, $d_C=0.051$, $C_1=39$ fF, $C_2=39$ fF, $C_{1g}=61$ fF, $C_{2g}=87$ fF, and $C_c=20$ fF where $E_J^i$ are the values at the zero flux points \cite{Kounalakis2018}. For $E_J^i\gg E_C^i$ as is the case for this device, Eq.~\ref{fullJac} may be approximated as Eq.~\ref{Jac} given the small contributions to the coupling of the second and third terms. When simulating the system, the full expressions were used.

\section{Extraction of Couplings}\label{SI-4-Couplings}
We begin with the full system Hamiltonian given in Sec. \ref{SI-4-Circuit}. Most terms are fast-rotating in the frame of the drives, but the choice of modulation frequency can selectively activate certain interactions. Whether modulating at the sum or difference frequency $\omega_m=\omega_A\pm\omega_B$, we retain the static terms which account for the frequencies, self-Kerr, and cross-Kerr coupling of the two transmons. Neglecting higher-order interactions with negligible effect and the small contributions of the sloshing mode, we are left with two Kerr-nonlinear oscillators
\begin{equation} 
\begin{aligned}
H_{DC}=\omega_Aa^\dagger a+\frac{\alpha_A}{2} a^\dagger a^\dagger a a+\omega_Bb^\dagger b+\frac{\alpha_B}{2}b^\dagger b^\dagger b b  + Va^\dagger a b^\dagger b
\end{aligned}.
\end{equation} 
When modulating at the difference and sum frequencies, we have in addition, the couplings given by
\begin{equation} 
\begin{aligned}
H_{\Delta}=(J_{AC} + J_{n_A}a^\dagger a+ J_{n_B}b^\dagger b+ J_{n_S}s^\dagger s)(a^\dagger b + ab^\dagger)
\end{aligned}
\end{equation} 
\begin{equation} 
\begin{aligned}
H_{\Sigma}=(J_{AC} + J_{n_A}a^\dagger a+ J_{n_B}b^\dagger b+ J_{n_S}s^\dagger s)(a^\dagger b^\dagger + ab)
\end{aligned}
\end{equation} 
which includes the occupation-dependent modifications of the single-photon hopping and two-mode squeezing interactions. In determining the total interaction strength under red and blue sideband modulation, we consider $\tilde{J}_{AC}=J_{AC}+J_{n_A}n_A+J_{n_B}n_B+J_{n_S}n_S$ where $n_i$ are the photon number expectation values of each mode, simply fitting to the total strength of the observed interaction $\tilde{J}_{AC}$. The expected values of $J_{n_i}n_i$ for the flux operation points in the main text are at most an order of magnitude lower than $J_{AC}$.

The presence of crosstalk between the flux ports could allow for the modulation of each of the two transmon SQUIDs, which would produce contributions to the interactions generated by the coupler alone \cite{Didier2018, Caldwell2018}. Thus, while $\tilde{J}_{AC}$ is the total strength of the hopping and squeezing interactions which we observe upon modulation of the flux signal incident on the coupler SQUID, the magnitude of the observed interaction is likely not entirely due to the coupler alone. In previous measurements on this device reported in Ref.~\cite{Kounalakis2018}, the DC flux crosstalk was found to be approximately 10\% between the flux ports of the coupler and those of each transmon.

\begin{table*}[t]
    \setlength\tabcolsep{6pt}
    \begin{tabular*}{0.66965\linewidth}{|c|c|c|c|}
    %\begin{tabular}{ |P{9cm}||P{2cm}|P{3.5cm}|P{3.5cm}|  }
    \hline
    \multicolumn{4}{|c|}{Extracted Parameters} \\
    \hline
    Name & Variable & RSB Value & BSB Value \\
    \hline
    Frequency of Transmon A   & $\omega_A/2\pi$ & \num[round-precision=3]{6.471618860087871} GHz &   \num[round-precision=3]{6.704480150101198} GHz\\
    Frequency of Transmon B  &  $\omega_B/2\pi$ & \num[round-precision=3]{4.712857972861132} GHz & 
    \num[round-precision=3]{5.573311878405953} GHz\\
    Anharmonicity of Transmon A   & $\alpha_{A}/2\pi$    & \num[round-precision=3]{-244.2019523816038} MHz &   \num[round-precision=3]{-241.25488872182362} MHz\\
    Anharmonicity of Transmon B  & $\alpha_{B}/2\pi$ & \num[round-precision=3]{-238.18342064988755} MHz   & \num[round-precision=3]{-236.37152440873214} MHz\\
    Linewidth of Measured Transmon  & $\kappa/2\pi$ & \num[round-precision=3]{2.4392566032625704} MHz   & \num[round-precision=3]{1.2986761172784522} MHz\\
    Flux Bias of Transmon A     &   $\Phi_A/\Phi_0$ & \num[round-precision=3]{0.11091972334275574} &  \num[round-precision=3]{0.016495661682120827}\\
    Flux Bias of Transmon B     &   $\Phi_B/\Phi_0$  & \num[round-precision=3]{0.4749955699148244} & \num[round-precision=3]{0.32022347574085774}\\
    Flux Bias of Coupler      &  $\Phi_{DC}/\Phi_0$  & \num[round-precision=3]{0.34858558795070177}   & \num[round-precision=3]{0.2145176538613559}\\
    AC Modulation Strength      &  $\Phi_{AC}$  & $\Phi_{DC}$/\num[round-precision=3]{17.531236521246974} & $\Phi_{DC}$/\num[round-precision=3]{47.708219663311006}\\
    Hopping / Squeezing & $\tilde{J}_{AC}/2\pi$  & \num[round-precision=3]{7.4615048344366794} MHz & \num[round-precision=3]{1.1313233699982654} MHz\\
    Cross-Kerr      & $V/2\pi$  & \num[round-precision=3]{-6.5428590015342545} MHz & \num[round-precision=3]{-9.158018288751824} MHz\\
    % Correlated Hopping / Squeezing & $\{J_{n_A}/2\pi$, $J_{n_B}/2\pi\}$   & \{\num[round-precision=3]{214.166278988}, \num[round-precision=3]{51.5736}\} kHz & \{\num[round-precision=3]{51.558}, \num[round-precision=3]{3.31565}\} kHz\\
    %  Cross-Kerr (Sloshing)  & $\{V_{AS}/2\pi$, $V_{BS}/2\pi\}$  & \{\num[round-precision=3]{-10.964648057400504}, \num[round-precision=3]{-14.815536645931983}\} MHz & \{\num[round-precision=3]{-13.515397}, \num[round-precision=3]{-16.132839}\} MHz\\
    %  Correlated Photon Hopping / Two Mode Squeezing (Sloshing) & $J_{n_S}/2\pi$   & \num[round-precision=3]{3.107229684347116} MHz & \num[round-precision=3]{0.565483} MHz\\
    \hline
    \end{tabular*}
    \caption{Parameters determined from spectroscopy measurements and extracted from fits to level repulsion and level attraction models for the data shown in Fig. \ref{RSB Modulation} and Fig. \ref{BSB Modulation}.}
    \label{Table 1}
\end{table*}

We first fit our measurements to an analytical equation following the method of Ref. \cite{Bernier2018}. For the case of red sideband modulation, we fit the real component of the level repulsion equation, which yields the system eigenfrequencies
\begin{equation} 
\begin{aligned}
    \omega_{\Delta}=\frac{\omega_A+\omega_B}{2}\pm\sqrt{(\frac{\omega_A-\omega_B}{2})^2+\tilde{J}_{AC}^2}
\label{omegadiff}
\end{aligned}
\end{equation} 
and under blue sideband modulation, the eigenfrequencies for level attraction are given by
\begin{equation} 
\begin{aligned}
    \omega_{\Sigma}=\frac{\omega_A+\omega_B}{2}\pm\sqrt{(\frac{\omega_A-\omega_B}{2})^2-\tilde{J}_{AC}^2}
\label{omegasum}
\end{aligned}.
\end{equation}

Following Ref. \cite{Bernier2018}, we can expand on Eq. \ref{omegasum} by including the self-Kerr and cross-Kerr terms to the system Hamiltonian. For oscillator A, these simply shift the frequency $\omega_A\rightarrow\omega_A+\alpha_An_A+Vn_B$ and for oscillator B $\omega_B\rightarrow\omega_B+\alpha_Bn_B+Vn_A$. Then, when the blue sideband modulation is applied at the appropriate frequency, we expect to observe regions of level attraction at frequencies $\omega_\Sigma$ shifted by the self-Kerr and cross-Kerr-nonlinearities dependent on the oscillator states. The secondary set of eigenfrequencies shown in Fig.~\ref{BSB Modulation} reflects this cross-Kerr shifted feature. Then, using Eq. \ref{omegadiff} and \ref{omegasum}, we obtain the coupling strengths shown in Table \ref{Table 1}.

When performing the numerical simulations with QuTiP, we obtain the expectation value of the photon annihilation operator for either mode for the system subject to a drive $H_d=\epsilon_d(a^\dagger e^{-i\omega_dt}+ae^{i\omega_dt})$ when measuring qubit A and similarly $H_d=\epsilon_d(b^\dagger e^{-i\omega_dt}+be^{i\omega_dt})$ for qubit B. The time evolution of the system is calculated under driving while sweeping the modulation frequency through either the red or blue sideband. The data for the full time-evolution is then used to calculate the expectation value once the system has reached a steady state, which is compared with the results of the fit to the analytical equation in Fig.~\ref{RSB Modulation} and Fig.~\ref{BSB Modulation}.

\end{document}